\documentclass[twocolumn,showpacs,preprintnumbers,amsmath,amssymb,aps]{revtex4}
\usepackage{graphicx}
\usepackage{dcolumn}
\usepackage{bm}
\begin{document}
\bibliographystyle{apsrev}


\title{Ground state of a confined Yukawa plasma}

\author{C.~Henning$^1$}
\author{H.~Baumgartner$^1$}
\author{A.~Piel$^3$}
\author{P.~Ludwig$^{1,2}$}
\author{V.~Golubnichiy$^1$}
\author{M.~Bonitz$^1$}
\email{bonitz@physik.uni-kiel.de}
\author{D.~Block$^3$}
\affiliation{$^1$Institut f\"ur Theoretische Physik und Astrophysik, Christian-Albrechts-Universit\"{a}t zu Kiel,
D-24098 Kiel, Germany}
\affiliation{$^2$Institut f\"ur Physik, Universit\"{a}t Rostock, Universit\"{a}tsplatz 3
D-18051 Rostock, Germany}
\affiliation{$^3$Institut f\"ur Experimentelle und Angewandte Physik, Christian-Albrechts-Universit\"{a}t zu Kiel,
D-24098 Kiel, Germany}

\pacs{52.27.Jt,52.27.Lw,05.20.Jj,52.27.Gr}

\date{\today}

\begin{abstract}
The ground state of an externally confined one-component Yukawa plasma is derived analytically. In particular, the radial density profile is computed. The results agree very well with computer simulations on three-dimensional spherical Coulomb crystals. We conclude in presenting an exact equation for the density distribution for a confinement potential of arbitrary geometry.
\end{abstract}
\maketitle

\section{Introduction}
Plasmas in external trapping potentials have been attracting increasing interest over the last few years in many fields, including trapped ions, e.g. \cite{itano,drewsen}, dusty plasmas, e.g. \cite{goree,zuzic,hayashi} and electrons and positrons in Penning traps, see e.g. \cite{dubin} for an overview. Among the main reasons is that, in these systems, it is relatively easy to realize strong correlation effects in charged particle systems. Probably the most spectacular manifestation of these effects is Coulomb liquid behavior and crystal formation which has been found in various geometries. In particular, the ion crystals and the recently observed spherical dust crystals or ``Coulomb balls'' \cite{arp04} have triggered intensive new experimental and theoretical work, e.g. \cite{totsuji05,ludwig-etal.05pre,bonitz-etal.prl06}. The shell structure of these crystals, including details of the shell radii and the particle distribution over the shells has been very well explained theoretically by a simple model involving an isotropic Yukawa-type pair repulsion and an harmonic external confinement potential \cite{bonitz-etal.prl06}.

Still, it remains an open question, how the average particle distribution inside the trap looks like, if it is the same as in the case of Coulomb interaction. It is well known that in a parabolic potential, particles interacting via the Coulomb potential  establish a radially constant density profile. Here, we extend this analysis to a plasma with Yukawa interaction by solving a variational problem for the ground state 
density (Sec. II). Then, in Sec. III we demonstrate that screening has a dramatic effect on the density profile giving rise to a parabolic decrease away from the trap center. There we demonstrate that the result for the density profile can be directly 
generalized to any anisotropic confinement potential.
While our analysis is based on a continuous plasma model on the mean-field level,
we find (Sec. IV), by comparison with molecular dynamics simulations, that the results apply also to spherical crystals with a shell structure.

\section{Ground state of a confined plasma} 
We consider a finite one-component plasma (OCP) containing $N$ identical particles with mass $m$ and charge $Q$ in an external potential $\Phi$ with pair interaction potential $V$ described by the hamiltonian
\begin{equation}
	H=\sum_{i=1}^N \left\{\frac{p_i^2}{2 m}+
	\Phi({\bf r}_i)\right\}+\frac{1}{2}\sum_{i\ne j}^NV({\bf r}_i-{\bf r}_j).
\label{eq:h}
\end{equation}
The classical ground state energy follows from Eq.~(\ref{eq:h}) for vanishing momenta and can be written as \cite{kraeft-etal.jp06a,dubin}
\begin{equation}\label{eq:energy}
	E[n]=\int d^3r\, u({\bf r}),
\end{equation}
with the potential energy density
\begin{equation}
	u({\bf r})=n({\bf r})\biggl\{\Phi({\bf r})+
	\frac{N-1}{2N}\int d^3r_2\,n({\bf r}_2)V(|{\bf r}-{\bf r}_2|)\biggr\},
\end{equation}
being a functional of the density profile $n({\bf r})$, and we neglected correlation contributions. The ground state corresponds to the minimum of the energy (\ref{eq:energy}) with respect to the density profile with the restrictions 
that the density is non-negative everywhere and reproduces the total particle number, i.e. 
\begin{equation}
	\int d^3r \,n({\bf r})=N. 
\label{eq:n}
\end{equation}

This gives rise to the variational problem
\begin{equation}\label{eq:varproblem0}
	0\stackrel{!}{=}\frac{\delta\tilde{E}[n,\mu]}{\delta n({\bf r})},
\end{equation}
where
\begin{equation}
	\tilde{E}[n,\mu]=E[n]+\mu\biggl\{N-\int d^3r\, n({\bf r})\biggr\},
\end{equation}
and we introduced a Lagrange multiplier $\mu$ (the chemical potential) to fulfil condition (\ref{eq:n}). The variation leads to 
\begin{equation}\label{eq:varproblem}
	\Phi({\bf r})-\mu+\frac{N-1}{N}\int d^3r'\,n({\bf r'})V(|{\bf r}-{\bf r'}|)=0,
\end{equation}
which holds at any point where the density is non-zero. Also, Eq.~(\ref{eq:varproblem}) is equivalent to vanishing of the total force on the particles separately at any space point ${\bf r}$, cf. Sec.~\ref{force_s}. 

Equation (\ref{eq:varproblem}) is completely general, applying to any pair interaction $V$ and confinement potentials of arbitrary form and symmetry, see Sec.~\ref{anyconf_s}.
Of particular interest is the case of an isotropic confinement $\Phi({\bf r})=\Phi(r)$, which leads to an isotropic density distribution $n({\bf r})=n(r)={\tilde n}(r)\Theta(R-r)$ the outer radius $R$ of which is being fixed by the normalization condition (\ref{eq:n}) which now becomes $\int_0^R dr\, r^2 {\tilde n}(r)=N/4\pi$.

\section{Density Profile of a Yukawa OCP}
We now consider the case of an isotropic Yukawa pair potential, $V(r)=\frac{Q^2}{r}e^{-\kappa r}$ which trivially includes the Coulomb case 
in the limit $\kappa \rightarrow 0$. Carrying out the angle integration in the interaction energy in Eq.~(\ref{eq:varproblem})  we obtain \cite{bonitz-dufty04}
\begin{multline}\label{eq:varequation}
	\Phi(r)-\mu=2\pi\frac{N-1}{N}\frac{Q^2}{\kappa r}\int_0^R
	 dr'\,r'\tilde{n}(r')\times\\
	\Bigl[e^{-\kappa(r+r')}-e^{-\kappa|r-r'|}\Bigr].
\end{multline}
This equation is the desired connection between the ground state density ${\tilde n}(r)$ of the Yukawa plasma with the external confinement $\Phi(r)$. This integral equation can be solved for the density by differentiating two times with respect to $r$ \cite{phi2} with the result
\begin{equation}\label{eq:nr}
   4\pi\frac{N-1}{N}\: Q^2\tilde{n}(r)=\frac{2\Phi'(r)}{r}+\Phi''(r)-\kappa^2\Phi+\kappa^2\mu.
\end{equation}
The yet unknown Lagrange multiplier can be obtained by inserting this explicit solution into Eq.~(\ref{eq:varequation}), which is then treated as an equation for $\mu$, with the result
\begin{equation}
  \mu=\Phi(R)+\frac{R\,\Phi'(R)}{1+\kappa R}.
\end{equation}

\subsection{Parabolic confinement potential}
For the frequently encountered case of a parabolic external potential, $\Phi(r)=\frac{\alpha}{2}r^2$, we obtain for the density from Eq.~(\ref{eq:nr})
\begin{equation}\label{eq:parabolicsolution}
	n(r)=\frac{\alpha N}{4\pi (N-1) Q^2}\Bigl(c-\frac{\kappa^2 r^2}{2}\Bigr)\Theta(R-r),
\end{equation}
where the constant $c$ is given by
\begin{equation}\label{eq:c}
	c=3+\frac{R^2\kappa^2}{2}\,\frac{3+\kappa R}{1+\kappa R}.
\end{equation}
Finally, the outer radius $R$ limiting the density profile is calculated from the normalization (\ref{eq:n}) with the result
\begin{multline}\label{eq:r}
	-15\frac{Q^2}{\alpha}(N-1)-15\frac{Q^2}{\alpha}\kappa (N-1)R\\
	+15 R^3+15\kappa R^4+6\kappa^2 R^5+\kappa^3 R^6=0.
\end{multline}
This equation has four complex and two real solutions, only one of which is non-negative and thus constitutes the unique proper result entering Eq.~(\ref{eq:c}). In the Coulomb limit, Eq.~(\ref{eq:parabolicsolution}) reduces to the familiar result of a step profile,
\begin{equation}\label{eq:rc}
  n_c(r)=\frac{3\alpha}{4\pi Q^2}\frac{N}{N-1}\Theta(R_c-r),
\end{equation}
where the outer radius is given by
\begin{equation}\label{rc}
  R_c=\sqrt[3]{\frac{Q^2(N-1)}{\alpha}}=r_0\sqrt[3]{\frac{N-1}{2}},
\end{equation}
which is fixed by the number of particles and the constant density, the latter being controlled by the curvature $\alpha$ of the potential. In the right part of Eq.~(\ref{rc}) we introduced 
the length scale $r_0=\sqrt[3]{2Q^2/\alpha}$, which is the stable distance of two charged particles in the absence of screening \cite{bonitz-etal.prl06} and which will be used below as the proper unit for lengths, screening parameter and density.
Note that Eq.~(\ref{rc}) holds also for a weakly screened Yukawa plasma with $\kappa R\ll 1$. 

\begin{figure}[h]
\includegraphics[height=6cm,clip=true]{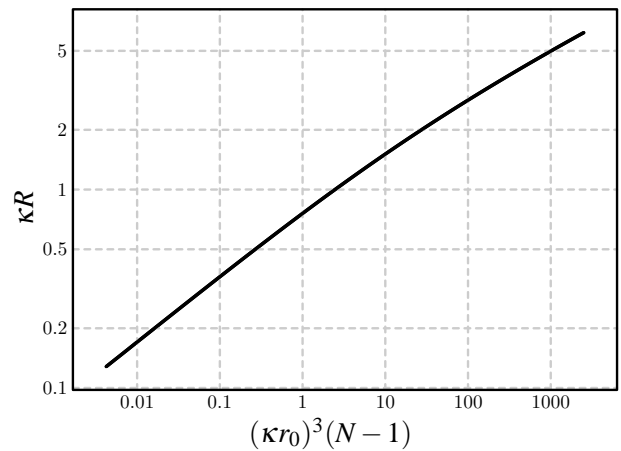}
\caption{Plasma cloud radius $R$ [positive real solution of Eq.~(\ref{eq:r})] for a parabolic confinement potential $\Phi(r)=\frac{\alpha}{2}r^2$ and Yukawa interaction 
with screening parameter $\kappa$.}
\label{fig1}
\end{figure}

In the other limiting case, $\kappa R \gg 1$, the radius has the asymptotics $\kappa R \approx [\frac{15}{2}(\kappa r_0)^3(N-1)]^{1/5}-1$. In general Eq.~(\ref{eq:r}) cannot be solved for $R$ explicitly. However, a general analytical result can be found by 
noting that all parameters entering Eq.~(\ref{eq:r}) combine into only two parameters, 
$x=(\kappa r_0)^3 (N-1)$ and $y=\kappa R$. Introducing these paramters into Eq.~(\ref{eq:r}), an explicit solution is found for the inverse function $x(y)$, which can be written as
\begin{equation}
	x(y)=\frac{2y^3}{15}\frac{y^3+6y^2+15y+15}{y+1}.
\end{equation}
Fig.~\ref{fig1} shows the result for the dimensionless radius $\kappa R=y$ of the plasma cloud, i.e. the solution of Eq.~(\ref{eq:r}) which for all values of 
$\kappa$ and $N$ is given by a single curve. 

\begin{figure}[h]
\includegraphics[height=6cm,clip=true]{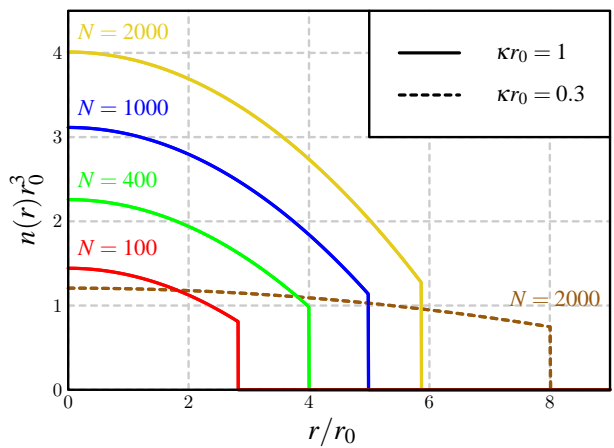}
\caption{(Color online) Radial density profile for a parabolic confinement potential $\Phi(r)=\alpha r^2/2$ and a constant screening parameter $\kappa r_0=1$ and four different particle numbers $N$ shown in the figure. For comparison, also the result for $\kappa r_0=0.3$ and $N=2000$ is shown by the dashed line.}
\label{fig2}
\end{figure}

With the result for $R$ the constant $c$, which is proportional to central density, 
can be computed from Eq.~(\ref{eq:c}), and the complete density profile, Eq.~(\ref{eq:parabolicsolution}), is found. The results are shown in Fig.~\ref{fig2} for four particle numbers between $N=100$ and $N=2000$. One clearly recognizes the 
inverted parabola which terminates in a finite density value, i.e. in a 
discontinuity, at $r=R$. 
With increasing $N$, the density increases continuously at every space point and, at the same time, extends to higher values $R$. Thereby the density profile retains its shape.

\begin{figure}[h]
\includegraphics[height=6cm,clip=true]{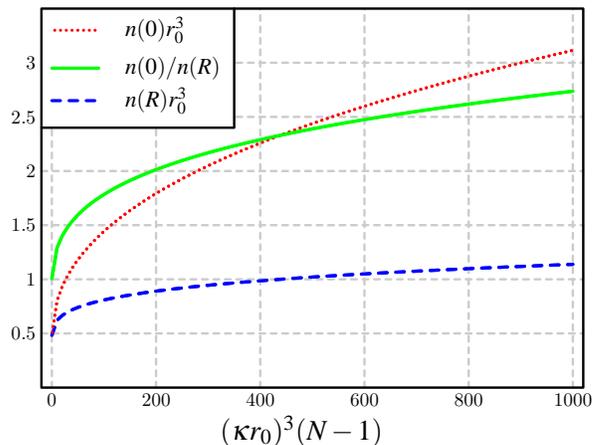}
\vspace{-0.5cm}
\caption{(Color online) Dependence of the central density $n(0)$ (red dotted line), density at the boundary, $n(R)$, (blue dashed line) and compression $n(0)/n(R)$ of the plasma (green full line) as a function of particle number and screening parameter.}
\label{fig:compression}
\end{figure}

On the other hand, when the plasma screening is increased, at constant $N$, the density profile changes dramatically, compare the two curves for $N=2000$. Increase of $\kappa$ leads to compression of the plasma: the radius $R$ decreases, and the absolute value of the density increases, most significantly in the center. This compressional behavior is shown in Fig.~\ref{fig:compression}, cf. the full green line showing the ratio of 
the inner to outer densities of the plasma. 

\begin{figure}[ht]
\includegraphics[height=6cm,clip=true]{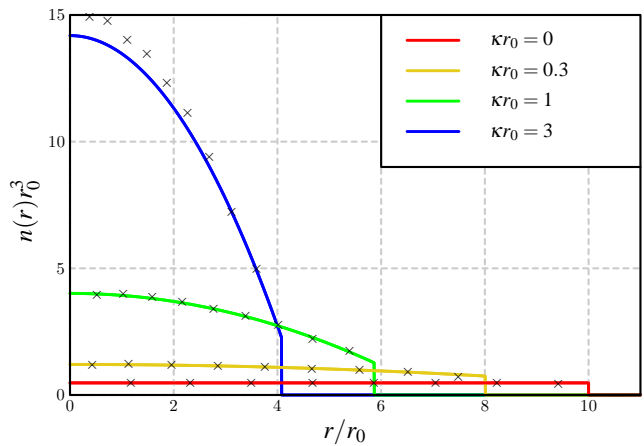}
\caption{(Color online) Radial density profile, solutions of Eq.~(\ref{eq:parabolicsolution}), of a three-dimensional plasma of $N=2000$ for four values of the screening parameter (lines), from bottom to top: $\kappa=0$ (red), $\kappa r_0=0.3$ (yellow), $\kappa r_0=1$ (green), $\kappa r_0=3$ (blue). Symbols denote molecular dynamics results of a plasma crystal for the same parameters where the average density at the positions of the shells is shown, for details see discussion in 
Sec.~\ref{yb_s}.}
\label{fig3}
\end{figure}

The dependence on $\kappa$ is analyzed more in detail in Fig.~\ref{fig3} below for a fixed particle number $N=2000$. In the case of Coulomb interaction, $\kappa=0$, we recover the constant density profile (\ref{eq:rc}). On the other hand, in the case of a screened potential, the density decays parabolically with increasing distance from the trap center, cf. Eq.~(\ref{eq:parabolicsolution}). Also, the density discontinuity at $r=R$ is softened compared to the Coulomb case, and the step height increases.

\subsection{Force equilibrium}\label{force_s}
Besides minimizing the total energy, cf. Eq.~(\ref{eq:varproblem}), the ground state density profile can be obtained from the condition of a local equilibrium of 
the total force (internal plus external ones) at each point where the density is non-zero. On the other hand, the shape of the radial 
density profile and its change with variation of $\kappa$ is directly related to a change of the force equilibrium. Here, we demonstrate this for the special case 
of a spherically symmetric confinement potential $\Phi(r)$.
The forces can be directly derived from Eq.~(\ref{eq:varequation}) by taking the gradient in radial direction
\begin{equation}\label{eq:forceeq}
\Phi'(r)=F_{<}(r)+F_{>}(r),
\end{equation}
which means that, for any spherical layer at a distance $r$ from the center, 
the external force $F_{\Phi}(r)=-\Phi'(r)$ which acts towards the center is balanced by the internal force due to the Yukawa repulsion between the particles. 
The internal force contains two parts where $F_{<}$
arises from the action of all particles inside the given layer, $r' \le r$, and acts 
outward, whereas $F_{>}$ results from the action of all particles located outside $r' \ge r$, and acts inward, 
\begin{align}\label{eq:forceequation}
	\begin{split}
	F_{<}(r) &= 4\pi\frac{N-1}{N} Q^2\frac{e^{-\kappa r}}{r}\left(1+\frac{1}{\kappa r}\right)\times\\
		 &\hspace{3cm}\int_0^r dr'\,r'\tilde{n}(r')\sinh(\kappa r')\notag,
	\end{split}\\
	\begin{split}
	F_{>}(r) &= 4\pi\frac{N-1}{N} Q^2\frac{1}{r}\left(-\cosh(\kappa r)+
	\frac{\sinh(\kappa r)}{\kappa r}\right)\times\\
	 &\hspace{3cm}\int_r^R dr'\,r'\tilde{n}(r')e^{-\kappa r'}\notag.
	\end{split}
\end{align}
This force balance can be used to obtain the ground state density profile. 
Alternatively, we can use the computed profile to analyze the two 
internal force contributions and their dependence on $\kappa$.

Consider first the limit of weak screening, $\kappa R\ll 1$. Then the forces approach the Coulomb case and, in the case of a constant density profile (\ref{eq:rc}),
\begin{subequations}
\begin{align}
F_{C,<}(r)&=\frac{N-1}{N}\frac{Q^2}{r^2}N_<=\alpha\:r,\\
F_{C,>}(r)&=0
\end{align}
\end{subequations}
with $N_<=n_c\,4\pi r^3/3$ being the particle number in the inner region. This means, 
the force is repulsive and increases linearly with $r$ and exactly compensates 
the linear external force $F_{\Phi}(r)=-\alpha r$ for all values $r<R$.

In the general case of finite screening the outer force $F_{>}(r)$ does not vanish, 
cf. Eq.~(\ref{eq:forceequation}). Since its direction is always towards the center, 
the force $F_{<}(r)$ has to increase simultaneously, in order to compensate the 
combined effect of $F_{\Phi}(r)$ and $F_{>}(r)$. This effect increases continuously 
with increasing $\kappa$ which is directly verified by evaluating the expressions 
in Eq.~(\ref{eq:forceequation}).

\subsection{Generalization to arbitrary confinement geometry}\label{anyconf_s}
The result for the density profile in an isotropic confinement, Eq.~(\ref{eq:nr}), can be easily extended to arbitrary geometry. For this purpose we use the text book result that the charge density corresponding to the Yukawa potential is $Q\delta({\bf r}) - Q \kappa^2e^{-\kappa r}/r$. This allows us to rewrite the Poisson equation as
\begin{equation}\label{eq:poisson}
	(\Delta-\kappa^2)\frac{e^{-\kappa r}}{r}=-4\pi\delta({\bf r}),
\end{equation}
showing that the Yukawa potential is the Green's function of Eq.~(\ref{eq:poisson}). This fact can be used in Eq. (\ref{eq:varproblem}) for the case of a confinement potential $\Phi$ of arbitrary geometry
\begin{equation}
	\Phi({\bf r})-\mu=-\frac{N-1}{N}\:Q^2\int d^3r'\,n({\bf r'})\frac{e^{-\kappa |{\bf r}-{\bf r'}|}}{|{\bf r}-{\bf r'}|},
\end{equation}
to get the explicit result for the density profile
\begin{equation}
	4\pi\frac{N-1}{N}\: Q^2n({\bf r})=\Delta\Phi({\bf r})-\kappa^2\Phi({\bf r})+\kappa^2\mu.
\end{equation}

\section{Density profile of confined Coulomb and Yukawa crystals}
So far we have considered the model of a continuous density distribution $n(r)$. 
On the other hand, the ground state of a confined spherically symmetric system of 
discrete point-like charged particles is known to have a shell structure as was 
demonstrated for dusty plasmas in Ref. \cite{arp04}. It is, therefore, of interest to 
verify if such a shell structure can be derived from our starting equation (\ref{eq:energy}) for the total energy and to 
compare our results to the radial density distribution in such Coulomb or Yukawa balls. 

\subsection{Derivation of a shell model for a trapped finite Yukawa plasma}
The concentric shells observed in spherical trapped Coulomb crystals have led to the proposal of simple analytical models, cf. e.g. \cite{hasse91,tsuruta93,totsuji05b,kraeft-etal.jp06a}. Such a model for a trapped one-component plasma is trivially derived from the total energy expression (\ref{eq:energy}) by inserting for the density the ansatz
\begin{equation}\label{eq:ns}
	n_s({\bf r})=n_s(r)=\sum_{\nu}^L \frac{N_{\nu}}{4\pi R_{\nu}^2}\delta(r-R_{\nu}),
\end{equation}
which describes $L$ concentric shells of zero thickness with $N_{\nu}$ particles on shell $\nu$ with radius $R_{\nu}$ and $\sum_{\nu=1}^L N_{\nu}+\zeta =N$, where $\zeta$ denotes the number of particles in the trap center (zero or one) \cite{tsuruta93,kraeft-etal.jp06a}.
As a result, we obtain for the total ground state energy of a Yukawa plasma in an isotropic general confinement potential $\Phi$
\begin{eqnarray}
&& E_s(N;\kappa)=\sum_{\nu=1}^L N_\nu\biggl\{\Phi(R_\nu)+Q^2
\frac{e^{-\kappa R_\nu}}{R_\nu} \times
\nonumber\\
&&\quad \biggl(\frac{\sinh(\kappa R_\nu)}{\kappa R_\nu}\frac{N_\nu-1}{2}+\zeta+\sum_{\mu<\nu}\frac{\sinh(\kappa R_\mu)}{\kappa R_\mu}N_\mu\biggr)\biggr\}.
\nonumber
\end{eqnarray}
This is essentially the Yukawa shell model of Totsuji et al., Ref.~\cite{totsuji05b} where, however, the finite size correction factor $(N_{\nu}-1)/N_{\nu}$ in the intrashell contribution and the term $\zeta$ are missing. In the Coulomb limit,  $\kappa\rightarrow 0$, the result simplifies with $e^{-\kappa R_{\nu}}\rightarrow 1$ and $\frac{\sin{\kappa R_{\nu}}}{\kappa R_{\nu}} \rightarrow 1$, and we immediately recover the Coulomb shell model of Hasse and Avilov \cite{hasse91} (plus the additional correction factor).

A further improvement is possible by including intrashell correlations \cite{tsuruta93}. The simplest model is obtained by replacing $N_{\nu}-1\rightarrow N_{\nu}-\epsilon(N)\sqrt{N_{\nu}}$, where $\epsilon$ is a fit parameter close to one which allows to achieve excellent agreement with the exact ground state \cite{kraeft-etal.jp06a}. An alternative way to include correlations was proposed by Ref. \cite{totsuji05b}.

\subsection{Comparison with simulation results for finite Yukawa crystals}\label{yb_s}
In order to compare the density profile $n(r)$ of our continuous model with the density of discrete spherical Yukawa crystals, we performed molecular dynamics simulations of the ground state of a large number of Coulomb balls, for details, see refs.~\cite{ludwig-etal.05pre, bonitz-etal.prl06}. As an example, the numerical results for a Coulomb ball with $N=2000$ which is large enough to exhibit macroscopic behavior \cite{totsuji02,schiffer02} are included in Fig.~\ref{fig3}.
The symbols denote the average particle density around each of the shells. The averaging was accomplished by substituting each particle by a small but finite sphere, so that a smooth radial density profile was obtained. 

With increasing $\kappa$ the crosses move towards the center confirming the compression of the Coulomb balls observed before \cite{bonitz-etal.prl06}. Obviously, the simulation results are very well reproduced by the analytical density profile (\ref{eq:parabolicsolution}) of a continuous plasma. But there are also small discrepancies in the central part which grow slowly with $\kappa$. One reason is that, for large $\kappa$, the width of the inner shells increases rapidly, making the comparison difficult. Another possible reason could be the effect of correlations.

\section{Summary and discussion}
In summary, we have presented a theoretical analysis of the ground state density profile of spatially confined one-component plasmas in dependence on the form of the pair interaction. An explicit result for the density 
profile for an arbitrary confinement potential has been derived. In particular, for an isotropic confinement, we have found that screening of the Coulomb interaction substantially modifies the radial density distribution. In contrast to a bare Coulomb interaction for which the density inside a parabolic external potential is constant, for a screened interaction, a quadratic decay away form the center is found.

Interestingly, while our results were derived for a continuous density distribution (a macroscopic system) and with neglect of binary correlations, our analytical results agree very well also with first-principle simulation results for strongly correlated 
Coulomb and Yukawa clusters containing several thousands of particles for screening 
paramters $\kappa r_0\le 1$. This agreement is by no means trivial and deserves further analysis. Further, it is very interesting to investigate the reason for the deviations 
at large values of the screening parameter, and analyze the effect of binary correlations on the density distributions is. These questions will be subject of forthcoming work.

\begin{acknowledgments}
This work is supported by the Deutsche Forschungsgemeinschaft via SFB-TR 24 grants A3, A5 and A7. 
\end{acknowledgments}


\begin{thebibliography}{26}
\bibitem{phi2} If $\Phi$ is not two times differentiable, the problem has to be solved separately in space regions where the second derivative exists.
\end{thebibliography}
\end{document}